\documentclass[twocolumn,prl,aps,flushbottom,showpacs]{revtex4-1}
\usepackage{xspace,amsmath,amsfonts,amsthm,amssymb,amsbsy,graphicx,color}
\begin{document}

\title{Absolute Dynamical Limit to Cooling Weakly-Coupled Quantum Systems} 

\author{X. Wang$^1$, Sai Vinjanampathy$^1$, Frederick W. Strauch$^2$, and Kurt Jacobs$^{1,3}$}

\affiliation{ 
$^1$Department of Physics, University of Massachusetts at Boston, Boston, MA 02125, USA \\$^2$Department of Physics, Williams College, Williamstown, MA 01267 \\
$^3$Hearne Institute for Theoretical Physics, Louisiana State University, Baton Rouge, LA 70803, USA
} 

\begin{abstract} 
Cooling of a quantum system is limited by the size of the control forces that are available (the ``speed'' of control). We consider the most general cooling process, albeit restricted to the regime in which the thermodynamics of the system is preserved (weak coupling). Within this regime, we further focus on the most useful control regime, in which a large cooling factor, and good ground-state cooling can be achieved. We present a control protocol for cooling, and give clear structural arguments, as well as strong numerical evidence, that this protocol is globally optimal. From this we obtain simple expressions for the limit to cooling that is imposed by the speed of control. 
\end{abstract}


\maketitle 

\newtheorem{theo}{Theorem} \newtheorem{lemma}{Lemma}

Preparing quantum systems in pure states is important for potential quantum technologies~\cite{OConnell10, Teufel11, Chan11, Verhagen12}, and this task is strongly linked to cooling to a (non-degenrate) ground-state: both require that all the entropy is extracted from the system. For this reason, there is presently a great deal of interest, experimental and theoretical, in cooling nano-mechanical resonators, and a number of cooling schemes of increasing effectiveness have been proposed~\cite{Mancini98, Wilson-Rae07, Marquardt07, Wang11}.  Since the forces used to implement cooling are always limited, the question of the maximum achievable ground-state population for a given maximum control force is important both fundamentally and practically. 
There are two structurally different regimes of cooling, that in which the dynamics of the thermal relaxation is preserved under the control (weak coupling), and that in which it is changed (strong coupling). Here we address the problem of optimal cooling in the former.

The complexity of the cooling problem --- and the closely related state-preparation problem --- is due to the interplay of both unitary and irreversible dynamics.
It is not usually possible to prove the optimality of control protocols for complex dynamical systems, and this has hampered progress in determining the fundamental limits of quantum control. 
We adopt a heuristic approach, and explore whether we can use the structure of the cooling problem to guess an optimal cooling protocol under dynamical constraints. We test the optimality of this protocol by comparison with those found using numerical optimization. We present very strong analytical and numerical evidence that we have explicitly constructed a globally optimal protocol, and thus determined the absolute dynamical limit to cooling in the dual regimes of weak coupling and good ground-state cooling. 

We consider the most general setting in which an $N$-dimensional ``target'' system can be cooled: the target system is coupled to a second, $M$-dimensional ``auxiliary'' system via an interaction Hamiltonian, $H_{\mbox{\scriptsize I}}$, whose eigenvalues we denote by $\hbar\lambda_j$ . This Hamiltonian, coupled with any trace-preserving operation on the auxiliary, implements the cooling process. The constraint we impose on the speed of control is that $| \lambda_j | \leq g, \forall j$, for some rate constant $g$. For a given experimental scenario, one calculates the eigenvalues $\lambda_j$ by i) determining the full Hamiltonian for the combined target and auxiliary systems; ii) removing all matrix elements that act on only one of the subsystems; iii) calculating the eigenvalues of the matrix that remains. 

Our results apply to all cooling schemes, including those involving continuous measurement via a probe system~\footnote{Since evaluating the performance of measurement-feedback requires averaging over the measurement process, the latter is fully accounted for within the trace-preserving operations on the auxiliary that we consider.}. However there is one class of schemes to which they are not immediately applicable: those in which the target is coupled directly to a near-continuum of states (e.g. a continuous measurement made directly on the target). This is because it is not yet clear whether our dynamical constraint is the right one from a practical point of view for this situation; the question of how it relates to the natural parameters for that case --- the ``damping rate'' or ``measurement strength'' --- is an interesting question for future work.    

To proceed we must chose some specific model of thermalization (environmental noise), and the Redfield master equation is the obvious choice~\cite{Breuer07}: it gives an accurate description of any weakly damped Markovian quantum system, and weakly damped systems are the most important for quantum technologies. (The Redfield master equation reduces to the standard quantum optical master equations for the Harmonic oscillator and two-level systems~\cite{Breuer07}.) We also start our analysis by allowing the auxiliary system to be ideal: it has large enough energy gaps that it has zero entropy at the ambient temperature (it is in its unique ground state), and is not itself subject to any damping or decoherence processes (we will see that if the auxiliary has zero entropy, damping it does not improve cooling). To a good approximation, present cooling methods for ions and nano-resonators are ideal in the first manner, but not the second. 

To cool the target, the control applied via the auxiliary must take the target out of equilibrium with its bath, and the bath therefore induces an irreversible relaxation of the system \textit{during} the cooling process. We must understand how the control and damping act together. Fortunately the regime in which cooling is most useful, and where present experiments operate, is that in which the target is cooled by a large factor $\textit{f} \gg 1$, where $f$ is the reduction in the average excitation number. This requires that the damping (thermalization) rate of the system, $\gamma$, satisfies $\gamma/g (\bar{n}+1) \ll 1$ (where $\bar{n}$ is defined below). The means that we can analyze this regime by considering the dynamics to first order in our small parameter, and this will allow us to obtain analytic results. If we define $\mathcal{D}(c)\rho \equiv (c^\dagger c \rho + \rho c^\dagger c)/2 - c \rho c^\dagger$ for some operator $c$, then the master equation that describes thermalization of a two-level system at temperature $T$ is given by $\dot{\rho} = - \gamma [(1 + \bar{n}) \mathcal{D}(\sigma) + \bar{n} \mathcal{D}(\sigma^\dagger)] \rho$, where $\bar{n} = \exp^{-\hbar\omega/kT}/(1 - \exp^{-\hbar\omega/kT})$. Here $\hbar \omega$ is the energy gap, $\sigma$ is the lowering operator, and the equilibrium population of the excited state is $P_{\mbox{\scriptsize T}} = \bar{n}/(1+2\bar{n})$.  For a harmonic oscillator the master equation is the same, but with the replacement $\sigma \rightarrow a$, $\omega$ is now the frequency of the oscillator, and $\bar{n}$ now gives the average number of photons/phonons at equilibrium. 

\begin{figure}[t]
\leavevmode\includegraphics[width=1\hsize]{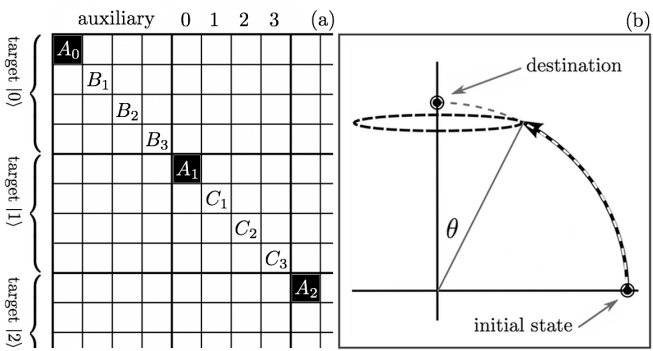}
\caption{(a) Depiction of the density matrix for the joint state of the target and auxiliary. The auxiliary states are the ``fast'' index, illustrated here with $M=4$. (b) A schematic depiction of the geometry of quantum dynamics, showing why rotations in the local subblocks do not change the angle-to-go for the cooling rotation. The ``great circle" that cools is the dashed arrow, and the dashed ellipse is the motion of a local rotation.} 
\label{fig1} 
\end{figure} 

Let us denote the energy levels of the target by $|m\rangle$, and those of the auxiliary by $|j\rangle_{\mbox{\scriptsize x}}$. If the auxiliary is prepared in the state $|0\rangle_{\mbox{\scriptsize x}}$, then the initial density matrix for the combined target and auxiliary is given by the matrix in Fig.~\ref{fig1}. Each subblock of this matrix is the full state-space of the auxiliary, and corresponds to a single state of the target. We see that the initial populations each appear in the elements labelled by $\mbox{A}x$, and are thus in different subblocks, and the task of cooling is that of transferring all the population to the subblock in the upper left-hand corner. The key observation we need is that cooling is a process of population transfer between orthogonal subspaces. Using the concept of majorization, it is readily shown that the largest possible population in any subspace occurs in the basis in which the density matrix is diagonal~\cite{Nielsen}, so we can expect that an optimal control protocol will keep the density matrix close to diagonal. 

Now that we have reduced the cooling problem to the transfer of population between orthogonal subspaces, insights from the geometry of quantum dynamics are now very useful~\cite{Anandan90, Margolus98, Carlini06}. The first of these is that given the constraint above, the fastest way to take any initial pure state to any other state is via a geodesic, the equivalent of a great circle in real vector spaces. The rotation angle along this great circle is determined by the inner product between the current and final states. The minimum time to get from any state $|1\rangle$ to an orthogonal state $|2\rangle$ is $\tau = \pi/(2g)$, and is achieved by the simple Hamiltonian $H = g (|1\rangle\langle 2| - |2\rangle\langle 1|)$. This tell us immediately that in the absence of damping (thermalization), the fastest way to perform the cooling operation is to rotate the states labelled by $\mbox{A}x$ in Fig.~\ref{fig1}, respectively to each of the unpopulated states labelled by $\mbox{B}x$, at this rate. 

A second fact, illustrated in Fig.~\ref{fig1}(b), is that local Hamiltonians of either subsystem cannot perform the population transfer needed to cool. Local Hamiltonians can only transform states to other states within specific subspaces -- we will call these the \textit{local spaces} of each system, and such population redistribution does not change the entropy of either system. Cooling therefore requires that we transfer population \textit{between} two or more of these local subspaces (e.g between the A and B subspaces). Further, from the dynamical geometry, if $H_{\mbox{\scriptsize I}}$ has partially rotated a state from one local space to another, no local Hamiltonian can change the angle (inner product) remaining between the rotated state and the destination B-state. Thus it cannot decrease the cooling time. 

Finally, by switching to the interaction picture, geometry reveals immediately that any action of a local Hamiltonian at any time can be undone, as far as the total population transfer is concerned, by varying $H_{\mbox{\scriptsize I}}$ with time: all we have to do is change the great circles upon which  $H_{\mbox{\scriptsize I}}$ rotates states, and this preserves the evolution of the total entropy of each system. Since we wish to obtain the optimal cooling over all possible time-dependent control protocols, this means that we can set the local Hamiltonians to zero, with no loss of generality.  With the above facts we can make the following general statement: 

\textit{Theorem:} Trace-preserving operations on an auxiliary system can reduce, but not increase, the maximum possible population transferrable to the ground state of the target system at any future time.

\textit{Proof} The population transferred to the ground state is the total population transferred between two orthogonal local subspaces, and this transfer is achieved most rapidly by a set of $90^\circ$ rotations on great circles connecting the two spaces. The action of a Hamiltonian local to the auxiliary cannot change the transfer rate. Further, i) any trace-preserving operation on the auxiliary is given by some unitary, $U$, acting on the auxiliary and an arbitrary third system, $\mbox{S}_3$, and ii) as far as the target is concerned,  $U$ is local in the sense above. Thus $U$ cannot increase the transfer rate, but it can reduce it: the interaction between the target and auxiliary, $H_{\mbox{\scriptsize I}}$, can rotate on all great circles that connect local subspaces of their joint space, but it does not have the same access to all subspaces of the  auxiliary and $\mbox{S}_3$. After the action of $U$, $H_{\mbox{\scriptsize I}}$ may no-longer be able to use the shortest paths for the transfer, and the transfer time will increase~\footnote{Note that the redundancy of a third system ($\mbox{S}_3$) is due to the fact that we allow the auxiliary to have zero entropy. If this were not the case, then $\mbox{S}_3$ could be valuable in purifying the auxiliary and improving cooling (e.g. algorithmic cooling~\cite{Schulman05}).}. $\Box$ 

\begin{figure}[t]
\leavevmode\includegraphics[width=1\hsize]{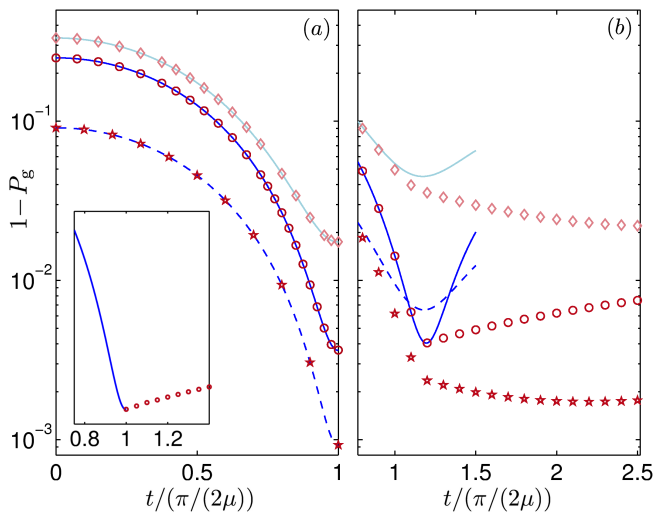}
\caption{(Color online) The performance of our conjectured optimal cooling protocol, compared against numerical optimization. The ground state population is $P_{\mbox{\scriptsize g}}$, and $1-P_{\mbox{\scriptsize g}}$ is shown vs. time for systems of size $N$, cooled by an auxiliary of size $M$, with control rate $g=1$, damping rate $\gamma=0.01$, and initial thermal factors $\bar{n}$.  (a) Cooling using an ideal (undamped) auxiliary, in which our protocol is optimal. Dark line: $(N,M)=(2,3)$, $\bar{n} = 0.5$; dashed line: $(N,M)=(4,4)$, $\bar{n} = 0.5$ light line: $(N,M)=(4,4)$, $\bar{n} = 0.1$; circles, squares, and diamonds: the corresponding results for numerical optimization. Inset: numerical optimization beyond the optimal cooling time (see text). (b) Cooling the same systems as in (a), but the auxiliary is strongly damped with rate $\kappa = 1$. Our protocol remains optimal for cooling a single qubit ($(N,M)=(2,3)$). } 
\label{fig2} 
\end{figure} 

Having established the Hamiltonian structure of the cooling problem, we now consider the continual thermalization of the target during the control process. We first present an argument that gives strong support for the following simple and rather remarkable statement:  all optimal cooling protocols will achieve the maximal ground-state population just prior to a time $\tau = \pi/(2g)$, and will do so only when the system starts in equilibrium.  This argument is as follows. The thermal master equation gives transition rates between the diagonal elements of $\rho$, and decay rates for all off-diagonal elements (in the energy basis).   As soon as the ground-state population rises above the equilibrium value, there is a net thermal transition rate out of the ground state to states orthogonal to it.  The population taken out of the ground state can be returned by transferring it to the auxiliary, which, as established above, will require a minimum time $\tau$.  During this time, a further amount of population, $P(\tau)$, will be taken out by thermalization. It is this amount of population that we will never be able to restrict to the ground state, since this population will again flow out by the time we have transferred it all back.  Furthermore, since population transfer is a rotation on the unit sphere, the rate of the increase of the ground-state population goes to zero as $t$ approaches $\tau$.  Since the exit rate from the ground state is nonzero at $t=\tau$, no matter how large $g /\gamma$, there will always be a time slightly prior to $\tau$ at which we lose by waiting longer. Finally, since the outward transition rate increases as the ground-state moves away from equilibrium, $P(\tau)$ will only be minimal when the ground-state \textit{starts} with its equilibrium value. This implies that the maximal cooling can only be obtained instantaneously; no steady-state cooling protocol can achieve this maximum. 
  
To arrive at our proposed optimal cooling Hamiltonian, we need to examine $\textit{where}$ thermalization places the population that leaves the ground state. For two-level systems and harmonic oscillators, thermalization involves transition rates only between adjacent energy levels, and so we focus on this here (the analysis generalizes easily). To first-order in $\gamma \tau = \pi\gamma/(2g)$, population from $\mbox{A}0$ goes to $\mbox{A}1$, and from $\mbox{B}x$ to $\mbox{C}x$, $\forall x$. During the transfer, the leakage contributions to the populations in $\mbox{A}1$ and $\mbox{C}x$ thus are first-order in $\gamma \tau$, and so much less than unity. Nevertheless, to obtain optimal cooling to first-order in $\gamma\bar{n}/g$ we must also optimally transfer these populations to the ground state. The population in $A1$ is already being optimally rotated to $\mbox{B}1$, and since the leakage population is small, and distributed over the great circle path, it is clear that the best option is simply to maintain this rotation: the presence of thermalization does not change the optimal protocol. The populations that appear in $\mbox{B}x$, $\forall x$, can be rotated to states in the B subspace that are not already destinations of existing rotations, if the auxiliary is large that additional B-states exist. This tells us two things. First, nothing can be done to retrieve the populations in $\mbox{B}$ states if the auxiliary is only as large as the target. Second, once the Hamiltonian is chosen to rotate all the $C$ states to the ground state, all the first-order leakage is handled optimally, and no further cooling is possible (to first-order). 

If the protocol above is indeed optimal, then given the structure of the thermal transition rates in Harmonic oscillators and single qubits, the maximum cooling can be achieved with an auxiliary dimension $M = 2N-1$, and the majority of the cooling with $M = N$. Further, the optimal interaction Hamiltonian, for an auxiliary with dimension $M$, would be $H_{\mbox{\scriptsize I}}^{\mbox{\scriptsize opt}} = G + G^\dagger$, where  
\begin{eqnarray}
  G  & = & g \!\!\!\!\!\!\!\!\!  \sum_{j=1}^{\min (M, N-1)} \!\!\!\!\!\!\!\!  |0, j\rangle \langle j, 0| + g \!\! \!\!\!\!\!\!\!\!\!\!\! \sum_{j=N}^{\min (M-1,2N-1)} \!\!\!\!\!\!\!\!\!\!\!\!  |0, j\rangle \langle 1, j\!\!-\!\!N\!\!+\!\!1|  , 
\end{eqnarray} 
and $|m,n\rangle$ means target state $|m\rangle$ and auxiliary state $|n\rangle$.  
  
Using the above cooling Hamiltonian we can now calculate explicitly the final ground-state population for a single qubit and a harmonic oscillator. To do this we use the linear version of the quantum-jump stochastic master equations that are equivalent to the thermal master equations~\cite{WisemanLinQ}. This allows us to exploit the fact that all the dynamics to first-order in $\gamma \tau$ is captured by trajectories containing no more than one jump. Further, we find that to first-order, the maximum ground-state population is reached at $t = \tau$ (the fact that the true maximum is slightly before $\tau$ is a second-order effect). For a single qubit, the minimum population left in the excited state is  
\begin{equation}
   P_{\mbox{\scriptsize min}} =  \frac{\pi\gamma}{4g} P_T \left(  \frac{1 -  P_T/4 }{1-2 P_T}  \right)   ,   \;\;\;\; \frac{\gamma}{g} \frac{(1-P_T)}{(1-2 P_T)} \ll 1 
\end{equation}
and $P_T$ is the excited-state population at the ambient temperature. Thus if $P_T$ is initially small, the fractional reduction in the excited-state population (the approximate ``cooling factor'') is  simply $\pi\gamma /(4g)$. 

For the harmonic oscillator, the minimum population that remains outside the ground state is 
\begin{equation}
   P_{\mbox{\scriptsize min}} =  \frac{\pi\gamma}{4g} \bar{n}  \left( 1 +  \bar{n}\frac{(3+\bar{n})}{4(1+\bar{n})^2} + \bar{n}^2\frac{(3+\bar{n})}{2(1+\bar{n})^2} \right) , 
\end{equation}
where $\bar{n}$ is the average number of phonons/photons in the target at the ambient temperature. When $\bar{n}$ is small, the ground state population at the ambient temperature is $1 - \bar{n}$, so the approximate cooling factor is, as one would expect, the same as for the two-level system. 

We now turn to the second part of our analysis, that of using numerical optimization to find optimal cooling protocols (we use the BFGS gradient search method~\cite{Nocedal06}). We perform these optimizations with a range of target and auxiliary dimensions. This analysis confirms, as far as it can, all of our assertions above: i) our protocol is optimal; ii) no more than $2N+1$ auxiliary states are required for cooling resonators; iii) the best cooling is obtained just prior to $t = \pi/(2g)$. Further, we find that up to unimportant transformations, the numerical optimization always finds our protocol, suggesting that it is unique. In Fig.~\ref{fig2}(a) we show results for a two-level target with a three-level auxiliary ($\bar{n} = 0.5$), and two cases of a target and auxiliary both with four levels, with $\bar{n} = 0.5$ and $\bar{n} = 0.1$ (The latter gives us results for a resonator.) These show that the optimization agrees with our protocol. The inset shows cooling using our protocol up to the optimal time (for $(N,M)=(2,3)$), and the circles give numerical optimization past this time, confirming that no better cooling is achieved for $t>\pi/(2g)$. We have also run the optimization for larger auxiliaries. All these results use $\gamma = 0.01$, $g=1$. 

All our analysis so far has assumed an ideal auxiliary system, since we are interested in the absolute limit to cooling. But nanoresonators are presently cooled via auxiliaries with significant damping, so finding optimal protocols with damped auxiliaries is an interesting question for future work. In Fig.~\ref{fig2}(b) we compare our protocol to numerical optimization, when the auxiliary is very strongly damped (damping rate $\kappa = g$).  Remarkably, for a two-level system our protocol appears to remain optimal, and the maximal cooling remains essentially the same. This is very interesting because this also applies to resonators in the low-temperature limit. For larger systems, we find that our cooling protocol is no longer optimal, and the optimized cooling is also degraded by the damping, as we would expect. 
 
\textit{Acknowledgements:} All the authors are partially supported by the NSF under Project No.  PHY-0902906, and all but FWS are partially supported by the NSF under PHY-1005571, and the ARO MURI grant W911NF-11-1-0268. 

\vspace{-7mm}
 
%\bibliography{report} 


\end{document}